\documentclass[twocolumn,showpacs,showkeys]{revtex4}
\usepackage{amsmath,graphicx}

\begin{document}

\title{Korean Family Name Distribution in the Past}
\author{Hoang Anh Tuan Kiet}
\affiliation{Department of Physics, BK21 Physics Research Division, and
Institute of Basic Science, Sungkyunkwan University, Suwon 440-746,
Republic of Korea}
\author{Seung Ki Baek}
\affiliation{Department of Physics, BK21 Physics Research Division, and
Institute of Basic Science, Sungkyunkwan University, Suwon 440-746,
Republic of Korea}
\author{Hawoong Jeong}
\affiliation{Department of Physics, Korea Advanced Institute of Science and
Technology, Daejeon 305-701, Republic of Korea}
\author{Beom Jun Kim}
\email[Corresponding author, E-mail: ]{beomjun@skku.edu}
\affiliation{Department of Physics, BK21 Physics Research Division, and
Institute of Basic Science, Sungkyunkwan University, Suwon 440-746,
Republic of Korea}

\begin{abstract}
We empirically study the genealogical trees of ten families for
about five centuries in Korea. Although each family tree contains
only the paternal part, the family names of women married to the
family have been recorded, which allows us to estimate roughly the
family name distributions for the past five hundred years.
Revealed is the fact that the unique Korean family name
distribution, characterized by a logarithmic form of the
cumulative distribution and an exponentially decaying rank-size
plot often called the Zipf plot, has remained unchanged for a
long time. We discuss the implications of our results within a
recently suggested theoretical model and compare them with
observations in other countries in which power-law forms are
abundantly found.
\end{abstract}

\pacs{87.23.Cc, 89.65.Cd}
%87.23.Cc Population dynamics and ecological pattern formation
%87.23.Ge Dynamics of social systems
%89.65.Cd Demographic studies
%89.65.Ef Social organizations; anthropology
%89.65.-s Social and economic systems
\keywords{Family name distribution, Rank-size plot, Zipf plot,
Population dynamics}

\maketitle

\section{Introduction}\label{sec:intro}

In the statistical-physics community, there has been intensive
research interest in the form and the origin of various
distributions observed in nature and in human societies
\cite{newman:powerlaw}. To name a few, the degree distributions of
many complex networks have been found to have power-law forms
\cite{network}, and the word frequency in literature, the sizes of
cities, the wealth distributions~\cite{jkps1}, the stock price
returns~\cite{jkps2}, and magnitude of earthquakes have also been
studied \cite{newman:powerlaw}. The family name distributions have
also been a popular research arena for statistical physicists, and
power-law forms have been observed in many countries
\cite{family:power,family:japan} with some exceptions
\cite{family:korea,family:baek,Yuan}.

Systems of family names have been developed in most countries in
order to make distinctions among families for various biological,
sociological, and economical reasons: For instances, marriages
among close family members have disastrous genetic effects, most
people want to keep the wealth of the family within the family,
and giving a social identity (you are one of us) to family members
has huge advantages in farming, construction, and forming a bigger
army. Once introduced, the inheritance of the family name in most
countries follows a very simple rule: children's family names are
from the father. However, the time evolution of the family name
distribution is not yet completely understood. A recent study
\cite{family:baek} has indicated the importance of the name
generation rate in explaining the very different family name
distributions observed across countries. In particular, two groups
of family name distributions have been found \cite{family:baek}:
Korea \cite{family:korea} and China \cite{Yuan} exhibit
exponentially decaying rank-size plots, or Zipf plots for the size
versus the rank of families, and the number of family names
increases logarithmically with population whereas for all other
countries where the empirical studies have been made, both follow
a power-law form.

\begin{table}
\caption{ Ten digitized family books, containing information on
women married to the family, used in the present
work~\cite{privacy}. Only women for whom their birth years and
family names were properly recorded are used in analysis ($N$ is
the number of such women in each data set). Women do not change
their family names after marriage in Korea. Here are also listed
the numbers of family names ($N_f$) and those considering the
regional origins ($N_r$). }
\begin{tabular*}{0.9\hsize}{@{\extracolsep{\fill}}ccccr}
\hline\hline
  data set  & year begun & $N$ & $N_f$ & $N_r$ \\
  \hline
  1 & 1513   & 104,366 & 165 & 2,668 \\
  2 & 1562   &  29,139 & 142 & 1,274 \\
  3 & 1439   &  17,911 & 121 &   923 \\
  4 & 1476   &  16,379 & 106 &   727 \\
  5 & 1698   &  15,445 & 125 &   915 \\
  6 & 1254   &  15,007 & 107 &   958 \\
  7 & 1475   &  11,526 & 112 &   736 \\
  8 & 1458   &   6,463 &  99 &   548 \\
  9 & 1752   &   3,500 &  89 &   390 \\
 10 & 1802   &   1,873 &  76 &   289 \\
\hline\hline\\
\end{tabular*}
\label{table:data}
\end{table}

In the present work, we use computerized data obtained from ten
family books, each of which had been updated and inherited within
a family from generation to generation. Each family book contained
paternal genealogical trees; in addition, the names of women
married to the family had been well recorded. In the analysis made
here, we only use the names of married women with their birth
years and their well-recorded family names (see Table 1). We then
take snapshots of the family name distributions in the past, which
reveals that the unique Korean name distribution has not changed
for at least five hundred years.

This paper is organized as follows: In Sec.~II, %\ref{sec:results}
we report our main empirical results and discuss in
Sec.~III %\ref{sec:region}
how different the observations are if we
include information on the regional origins of family names.
Comparisons with other countries and the implications of our
results in a recently developed theoretical approach are discussed
in Sec.~IV.%\ref{sec:discuss}

\section{Korean family name distribution in the past}
\label{sec:results}

In Korean tradition, it is extremely unlikely that any Korean
changes his (her) family name, which explains the existence of the
small number of family names found in Korea (288 family names
\cite{family:korea} in comparison to 132,000 names in Japan
\cite{family:japan}). Such a small number of family names in Korea
inevitably makes some families huge. For example, the biggest
family name, Kim, covers more than 20\% of total population in
Korea. Consequently, members in the huge-sized family cannot have
a social group identity as a family, and classifications into
subgroups on finer scales have naturally been developed. More
specifically, the regional origin of the family combined with the
family name has been used to distinguish one family from the other
(see Ref.~7 for more discussion); one of the
authors of the present paper belongs to the sub-family group ``Baek
from Suwon,'' with Suwon being the regional origin of his ancestor.
We, in this Section, report our results for the case when the
regional origins are disregarded; i.e., families with the same
name but with different regional origins are taken as one family.
Observations made for other case with the regional origin taken
into account are presented in the next Section.

\begin{figure}
\includegraphics[width=0.45\textwidth]{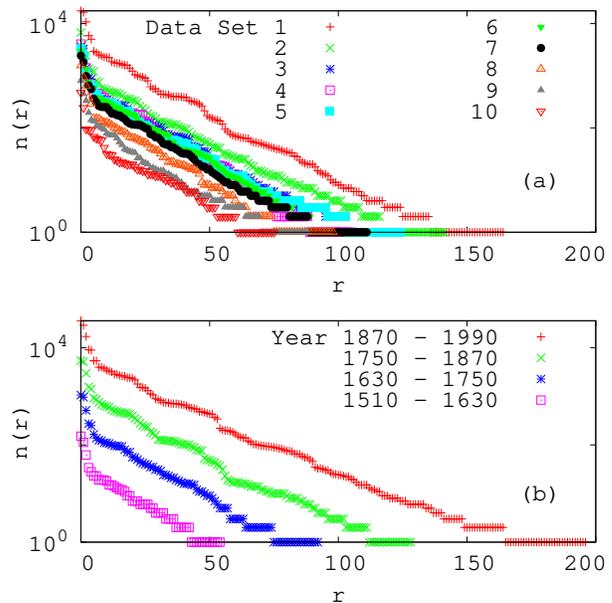}
\caption{(Color-online) (a) The size $n(r)$ versus the rank $r$ of
families (Zipf plot) taken from the 10 different data sets in
Table 1. (b) All data sets in Table 1 are merged and then split
according to birth years. It is clearly seen that (a) each data
set exhibits qualitatively the same exponentially decaying
rank-size plot and (b) the overall feature has not changed for
about five hundred years. } \label{fig:zipf}
\end{figure}

In Figure \ref{fig:zipf}, we display the rank-size plot
%\footnote{ Throughout the present work, we use the term 'the
%\textbf{rank-size plot}' to mean the so-called Zipf's plot in
%which the size and the rank are displayed in the vertical and the
%horizontal axes, respectively.}
for the number $n(r)$ of members, i.e., the size of the family,
versus the rank $r$ of the family. For example, if a data point
exists in the plot at ($r=2$, $n=10^3$), it means that the second
biggest family name has $10^3$ members within it. It is remarkable
to see in Figure \ref{fig:zipf}(a) that all data sets, small and
big, in Table 1 unanimously exhibit an exponential decay over a
broad range of $r$. Although the total size of available data sets
can cover only a tiny fraction of the whole population within
Korea for five hundred years, we strongly believe that the
observation of the same exponentially  decaying rank-size plot for
all ten data sets can hardly be a coincidence. In Figure
\ref{fig:zipf}(b), we show the rank-size plot when all ten data
sets are merged and then split into four different time periods,
which confirms that the exponentially decaying rank-size plot has
not changed for at least five hundred years in Korean
history. Consequently, we suggest that the already found
exponential rank-size plot in Ref.~7 from the
census results for the years 1985 and 2000 has been a unique
characteristic of the Korean family name distribution for a very
long time since the family name system was introduced at least
about two thousand years ago. Although limited by the available
size of the family name information, it is also interesting to
recognize that our method can provide a snapshot of the family
name distribution in the past at any given time span.

Even if a rank-size plot looks similar to another as a whole, the
actual order of frequent family names may be different for each
data set. Suppose that we pick a data set and make a list of
family names therein in a descending order of their family sizes.
Then, the rank will naturally increase one by one along this list.
In a different data set, however, the same list of family names
will have a different order of ranks, in general. For example, we
present Table~\ref{table:tau_example} where the order becomes
simply reversed. Denoting this new order as $\{r_i\}$ with $i=1
\ldots 5$, let us count $P$, the number of pairs $(r_i, r_j)$ such
that $i<j$ and $r_i<r_j$. One can easily see the answer to be zero
because the rank is only decreasing as we move to the right. On
the other hand, if the new order had been identical with the
original one, $P$ would have amounted to $s(s-1)/2$ with the
number of elements of $s=5$. Therefore the statistics $P$ may be
an indicator for the degree of rank correlation within the range
between zero and $s(s-1)/2$.
\begin{table}
\caption{An example of the rank correlation between two data sets
denoted as \#1 and \#2.}
\begin{tabular*}{0.9\hsize}{@{\extracolsep{\fill}}rccccr}
\hline\hline
family name & A & B & C & D & E \\\hline
ranks in \#1 & 1 & 2 & 3 & 4 & 5 \\
\#2 & 5 & 4 & 3 & 2 & 1 \\
\hline\hline
\end{tabular*}
\label{table:tau_example}
\end{table}
Normalizing $P$ into $[-1,1]$ gives the {\em
Kendall tau rank correlation coefficient} as follows~\cite{Abdi}:
\begin{equation*}
\tau = \frac{2P}{s(s-1)/2} - 1.
\end{equation*}
As discussed above, if the orders of ranks are unchanged across
two data sets, $\tau$ equals unity while $\tau=-1$ for a perfectly
reversed order. If rankings are independently given, $\tau$ will
approach zero on average. Table~\ref{table:tau} displays the rank
correlations between the data sets in Table~\ref{table:data}. The
correlation between two different data sets is estimated as
$\bar\tau = 0.73 \pm 0.06$, showing the significantly high
correspondence among one another.

\begin{table*}
\caption{The Kendall tau rank correlation coefficients between the data
sets used in the present work.
Each element at the row $i$ and the column $j$ indicates $\tau(i,j)$, the
Kendall tau rank correlation coefficient between the data sets $i$ and
$j$. Note that $\tau(i,j)=\tau(j,i)$, and that we consider only the family
names commonly appearing in both of the data sets under comparison.}
\begin{tabular*}{0.9\hsize}{@{\extracolsep{\fill}}c|cccccccccc}
\hline\hline
data set & 1 & 2 & 3 & 4 & 5 & 6 & 7 & 8 & 9 & 10 \\\hline
 1 & 1.00 & 0.78 & 0.73 & 0.84 & 0.77 & 0.77 & 0.64 & 0.82 & 0.79 & 0.77 \\
 2 &      & 1.00 & 0.75 & 0.76 & 0.74 & 0.70 & 0.61 & 0.70 & 0.66 & 0.70 \\
 3 &      &      & 1.00 & 0.71 & 0.75 & 0.73 & 0.67 & 0.71 & 0.73 & 0.74 \\
 4 &      &      &      & 1.00 & 0.76 & 0.75 & 0.58 & 0.80 & 0.74 & 0.77 \\
 5 &      &      &      &      & 1.00 & 0.70 & 0.62 & 0.80 & 0.73 & 0.72 \\
 6 &      &      &      &      &      & 1.00 & 0.63 & 0.75 & 0.78 & 0.81 \\
 7 &      &      &      &      &      &      & 1.00 & 0.61 & 0.65 & 0.65 \\
 8 &      &      &      &      &      &      &      & 1.00 & 0.78 & 0.77 \\
 9 &      &      &      &      &      &      &      &      & 1.00 & 0.75 \\
10 &      &      &      &      &      &      &      &      &      & 1.00 \\
\hline\hline
\end{tabular*}
\label{table:tau}
\end{table*}

The functional form of the rank-size plot is, of course, closely
related with the family name distribution function $P(k)$, which
measures the frequency of the families of size $k$
\cite{family:korea}. In particular, the exponential rank-size plot
leads to the functional form $P(k) \sim k^{-1}$, which yields the
logarithmic cumulative distribution function $P_{\rm cum}(k) \sim
- \ln k$. With the assumption of uniform sampling as discussed in
Ref.~7, the number $N_f$ of family names in a
population of size $N$ is expected to increase logarithmically,
i.e., $N_f \sim \ln N$. In comparison, for many other countries,
the observed power-law behavior of $P(k) \sim k^{-\gamma}$ gives
$P_{\rm cum}(k) \sim k^{1 - \gamma}$, and $N_f \sim N^{\gamma-1}$
is expected. Consequently, the exponential rank-size plot in Figure
\ref{fig:zipf} is expected to explain the logarithmic behavior in
the number of family names, i.e., $N_f \sim \ln N$, which is
closely related with such a small number of family names in Korea
in comparison to other countries~\cite{family:baek}.

\begin{figure}
\includegraphics[width=0.45\textwidth]{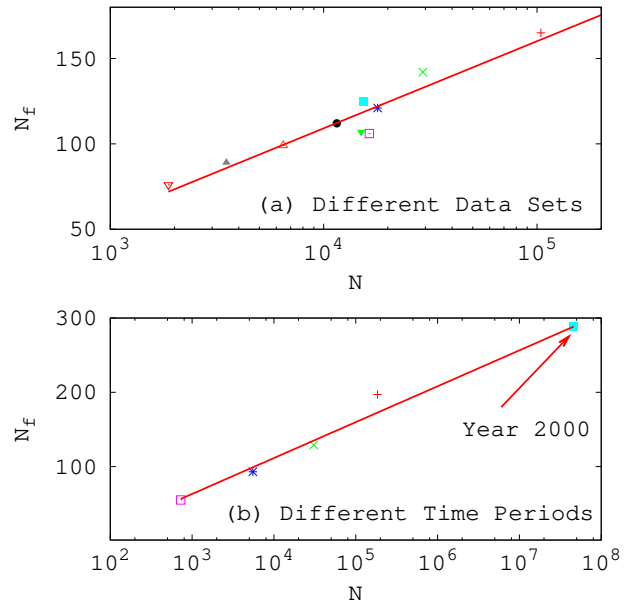}
\caption{(Color online) The number $N_f$ of different family names
without regional origins versus the population size $N$ (a) for
each date set in Table 1 and (b) for given time periods, in
parallel to Figs. \ref{fig:zipf}(a) and (b), respectively. Symbols
in (a) and (b) correspond to the ones in Figure \ref{fig:zipf}. It
is shown clearly that $N_f \sim \ln N$ in (a). In (b), the actual
result from the census in year 2000 is in accord with the
logarithmic form. } \label{fig:NfN}
\end{figure}

In order to validate the relation between the rank-size plot and
$N_f(N)$, we display in Figure \ref{fig:NfN} the results
corresponding to Figure \ref{fig:zipf}. In Figure \ref{fig:NfN}(a),
$N_f$ and $N$ are taken from data sets in Table 1 with symbols
corresponding to those in Figure \ref{fig:zipf}(a), and the overall
feature is well described by $N_f \sim \ln N$, as expected. In
Figure \ref{fig:NfN}(b), we again use the time-split data as in Figure
\ref{fig:zipf}(b), which is in accord with the logarithmic form,
with the point for the year 2000 included. We again emphasize that
our observation is in a sharp contrast to the finding in Japan
\cite{family:japan}, where $N_f$ has been found to increase
algebraically with $N$, not logarithmically.

\section{Family name distribution with regional origins}
\label{sec:region}

\begin{figure}
\begin{center}
 \includegraphics[width=0.45\textwidth]{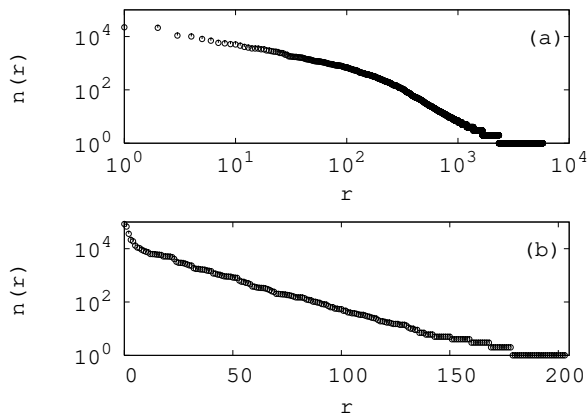}
  \caption{Rank-size plots (a) with and (b) without regional
  origins. The rank-size plot without regional origins in (b)
shows a very clear exponential decay as was observed in Figure
\ref{fig:zipf}, while it is much broader in (a) when the regional
origins of family names are taken into account. All women's names,
regardless of the existence of the birth-year information, are
included for better statistics. }
  \label{fig:notime}
\end{center}
\end{figure}

We next study the family name distribution in the past with the
regional origins taken into account. In other words, we now
distinguish ``Kim from Gimhae,'' to which one of the authors of the
present paper belongs, from ``Kim from Gyeongju,'' and thus count
the two as different family names. We believe that findings in
Sec.~II %\ref{sec:results} 
confirm the robustness of the family name
distribution for a long time; accordingly, in order to make the
sample size bigger, we include all women, regardless of the
presence or absence of the birth-year information. In Figure~\ref{fig:notime},
we summarize our results for the rank-size plot
(a) with regional origins taken into account and (b) with regional
origins disregarded, computed from all 420,719 women in comparison
to 221,609 in Table 1 and in Sec.~II. %\ref{sec:results}.
It is clearly
seen that the rank-size plot again shows an exponential decay form
whereas the rank-size plot when the regional origins are included
becomes very broad, presumably in a power-law form in the large
$r$ region. The total numbers of names with and without regional
origins are 5,788 and 204, respectively.
%giving the rough estimate
%that on average each family name has about 30 different
%regional origins.

\section{Discussion}
\label{sec:discuss}

Recently, different family name distributions across countries
have been explained as originating from different name generation
behaviors\textbf{\cite{family:baek}}.  In detail, from the use of
the master equation approach, we have shown that in a society
where new family names are rarely invented, the rank-size plot is
expected to be of an exponential form. On the other hand, if the
new name generation rate is proportional to the population size,
the rank-size plot is of the power-law form. We in this work have
shown that in the past five hundred years in Korea, the
exponential rank-size plot has been unchanged, which is closely
related with such a small number of family names in Korea
according to Ref.~7.

In summary, we have used computerized family genealogical tree
information for ten families to extract the birth years and the
names of women who were married to each family. Clearly shown is
the robustness of the family name distribution in Korea for the
past five hundred years, at least. Particularly, the rank-size
plot has been shown to be of an exponential form, which confirms
the finding in Ref.~7 made from the results of
the recent census. This is in itself an interesting surprise
because we expected that in Korean history, families in a higher
social class must have chosen women in equally influential
families, distorting the women's family name distributions.
Additionally, our empirical explorations have also indicated that
the family name distributions with and without regional origin
information of married women are very different: The latter takes
the exponential form, and the former has a much broader decay
form, in close resemblance to the family name distributions in
other countries.

\acknowledgments

We thank N-Korean (www.n-korean.com) for providing us data from
Korean family books. This work was supported by a Korea Research
Foundation Grant funded by the Korean Government (MOEHRD, Basic
Research Promotion Fund) with Grant No.
KRF-2006-211-C00010 (H.A.T.K.), KRF-2005-005-J11903 (S.K.B.), and
KRF-2006-312-C00548 (B.J.K.). H.J. was supported by the Korea
Science and Engineering Foundation with Grant No.
R01-2005-000-1112-0.


\begin{thebibliography}{00}
\bibitem{privacy} In order to protect the privacy of individuals, only the family name
field from the family book has been provided for data analysis.
\end{thebibliography}
\end{document}